# Near Resonant Spatial Images of Confined Bose-Einstein Condensates in the '4D' Magnetic Bottle.

Lene Vestergaard Hau[1,2], B.D. Busch[1,2], Chien Liu[1], Zachary Dutton[1,2], Michael M. Burns[1], and J.A. Golovchenko[1,2]

[1]Rowland Institute for Science, 100 Edwin H. Land Boulevard, Cambridge, MA 02142.
[2]Department of Physics, Harvard University, Cambridge, MA 02138.

Abstract.
We present quantitative measurements of the spatial density profile of Bose-Einstein condensates of sodium atoms confined in a new '4D' magnetic bottle. The condensates are imaged in transmission with near resonant laser light. We demonstrate that the Thomas-Fermi surface of a condensate can be determined to better than 1%. More generally, we obtain excellent agreement with mean-field theory. We conclude that precision measurements of atomic scattering lengths and interactions between phase separated cold atoms in a harmonic trap can be measured with high precision using this method.

Recently Bose-Einstein condensates have been created from dilute, ultra-cold atomic clouds of Rb, Li, and Na [1-5] through a combination of laser [6] and evaporative cooling [7]. Evidence for condensation in refs 1 and 3-5 rely on time of flight measurements on atomic clouds after release from the magnetic traps in which they are initially confined; valuable information on condensate dynamics has been obtained from studying such release data [8]. Alternatively, it is possible to probe confined condensates directly without the transformations associated with release processes. This has been done with dark-field and phase contrast imaging [9-11]. In this paper we describe such a capability obtained with near resonant absorption imaging in a BEC setup based on a new '4D' magnetic bottle in which we routinely create million atom condensates of sodium atoms. We show quantitative *in-situ* spatial images of the condensate surface region and perform detailed comparisons of density profile measurements on pure condensates (no visible non-condensate component) to ground state mean field calculations.

These condensates, confined in a harmonic trap and with large numbers of atoms (Thomas-Fermi limit [12]), have sharply defined boundaries that can be determined with high precision with near resonant imaging. Combined with an independent determination of the number of condensate atoms from release measurements, this forms the basis for a precise determination of the scattering length for condensate atoms.

The apparatus used in our experiments is illustrated in Fig. 1. A 'candlestick' atomic beam source [13] injects a beam of $4\times10^{16}$ sodium atoms/sec/sterad into a one meter long Zeeman slower [14]. Our version of the Zeeman slower incorporates a zero crossing in the magnetic field that varies from +350 to -350 gauss from entrance to exit. All laser beams in the setup are derived from a single dye laser and tuned with acousto-optic and electro-optic modulators.

The Zeeman slower produces a 50 m/s atomic beam that continuously loads a dark spot, magneto optic trap (MOT) [15]. The MOT is created by three pairs of 1" diameter, 10 mW/cm$^2$, circularly polarized, counter-propagating laser beams detuned 20 MHz below the F=2→3 resonance. Anti-Helmholtz coils provide a 10 gauss/cm field gradient. After 2 seconds of accumulation, the MOT contains $6 \times 10^9$ Doppler cooled atoms at a density of $6 \times 10^{11}/cm^3$ and a temperature of 250 µK.

After the MOT is filled, a mechanical shutter isolates the candlestick source from the confined atomic cloud which is then polarization gradient cooled [16] to 50 µK within a few milliseconds. Finally the atoms are pumped into the F=1 ground state and the laser beams are turned off.

All magnetic fields are created by electromagnets outside of the vacuum chamber without the aid of permeable cores. The fields can thereby be precisely controlled by programmed current supplies and, most importantly, switched on and off in milliseconds. Rapid switching of magnetic fields is effected with specially designed FET banks that maximize the switching rates through appropriate programming of the voltage applied to the coils.

The next stage of temperature reduction and density enhancement is provided by evaporative cooling in a newly designed '4D' magnetic bottle, so named because the shape of each of the four coils needed to create the confining field for spin aligned atoms resembles the letter 'D'. Fig. 1(b) shows the configuration of these coils. Electrical currents of up to 895 amperes are passed through each water cooled 'D'. At this level the system dissipates a total power of 7.5 kW. The '4D' bottle creates a magnetic field structure similar to that found in Ioffe traps [17]. Detailed design considerations will be presented elsewhere.

The field obtained from these magnets has a z component (see Fig. 1(b)) that varies parabolically along the z direction with a curvature of 128 gauss/cm$^2$ (at 895 A) and transverse components with a linear variation in the transverse directions with gradients of 223 gauss/cm. The figure of merit for the efficiency of electromagnets is the ratio of field gradient to electric power dissipation and the quoted numbers for the '4D' magnet make it the most efficient room temperature electromagnetic atom trap reported to date.

The quoted field parameters are limited by the maximum current output from our power supplies. Each coil consists of eight layers of machined copper D's, electrically connected in series but water cooled in parallel. This makes the cooling very efficient and five times larger currents could easily be applied.

In addition to the 'D' magnets, Helmholtz coils are used to create a tunable bias field (maximum 235 gauss) in the z direction. The total B field has a minimum but no zero crossing which eliminates trap loss caused by Majorana spin flip transitions. The value of the bias field controls the confinement that an atom will experience in the transverse directions. Confinement along z is controlled entirely by the 'D' magnets.

After polarization gradient cooling the atom cloud is quickly captured by energizing the '4D' trap with the bias field adjusted to minimize heating and density loss in the transfer process [18]. The atom cloud is then compressed adiabatically in the transverse directions, thereby increasing the atomic density and collision rate for efficient rf evaporative cooling which is the next step in the process. With the trap parameters held fixed, a transverse high frequency magnetic field is excited by passing 100 mA of rf current through coils placed around the confined atomic cloud. The frequency is scanned from 32 MHz down to approximately 1 MHz in 38 seconds in a schedule calculated from a model based on Ref. 19. The final rf frequency determines the final temperature, number of atoms and condensate fraction of the cloud. To minimize trap loss, chamber pressure is kept below $1 \times 10^{-11}$ torr.

In this report we are concerned with demonstrating that detailed quantitative spatial information for condensates *in* the trap can be obtained by near resonant absorption imaging. A few images of trapped condensates have been reported in the literature, but we are not aware of any demonstrations where the statistics [9-11] or resolution [20] have supported a detailed comparison of the spatial density profile of a trapped condensate with mean field theory. (For example, compare the boundary regions of the condensates in Figs 2 and 3 of this paper to Fig. 3 in ref. 11 obtained with phase-contrast imaging).

There have been fears in the community that a near resonant light beam passing through a condensate will give rise to such strong refractive effects that image reconstruction is impossible [9,10], and that heating effects associated with strong absorption by the condensate might distort the cloud during the imaging process. Both our theoretical studies and experimental results show otherwise.

To study the refraction problem we used the eikonal approximation [21] to estimate the absorption and deflection of an initially parallel bundle of optical rays as they pass through a condensate. In the real and imaginary parts of the dielectric function we include the oscillator strengths and resonant frequencies for all the atomic energy levels that are coupled by the incident laser probe beam to the ground state of the atom. The spatial dependence of the dielectric function is determined by the local density of atoms and local field effects (Lorentz-Lorenz corrections) are included. In the eikonal approach, the optical rays are propagated through the condensate and the final ray position, angle, phase and field strength are recorded in a plane directly behind the condensate. This plane is the object plane of the imaging optics which is illustrated in Fig. 1(a). The wavefield in the object plane is then propagated through the optics by Fourier techniques to the image plane [22, 23].

The results for condensate sizes and densities described in this paper indicate that for sufficient probe beam detunings ($\approx$40 MHz) and small enough f number optics, excellent images are obtained [23]. When a ray encounters index gradients large enough to cause deflection on our object plane or out of our f5 optics, it is already eliminated by absorption, leading to no loss of information (contrary to what was stated in [9-10]).

To image the condensates, a CCD camera is placed in the image plane of the optical system (Fig. 1(a)). The first two lenses of the imaging system (2 inch diameter achromatic doublets with focal lengths of 250 mm) create a 1:1 image just in front of the last lens (a 55 mm focal length macro lens) which magnifies the image on the CCD camera by a factor of 5. Experimental images were obtained from a 10 µs exposure to a 1 mW/cm$^2$ probe laser beam tunable ±45 MHz around the F=2→3 resonance frequency. The atoms were pumped for 10 µs prior to probing to the F=2 ground state

The problem of heating due to probe and pump laser beams was addressed empirically. For the probe and pump times chosen, no change in condensate profile resulted by increasing either time by a factor of two. Note that Fig. 3 is obtained from images of three different condensates with probe times ranging from 10 to 20 µs.

The width of each CCD camera pixel corresponds to 2 µm at the position of the condensate, which is about 1/5 of the spatial resolution of the optics (defined as the point resolution function for coherent light, calculated by taking lens aberrations into account) [24]. To compensate for spatial nonuniformities in the probe beam, the transmission profile is normalized to the probe transmission with no atoms present.

Fig. 2 shows the results of such imaging experiments performed on three differently sized condensates in the '4D' bottle. The probe detuning is -35 MHz and the bias field is 1.5 gauss. In each case we continue the evaporation process until we are left with a 'pure' condensate (no visible thermal cloud). The condensate profiles are shown along the z direction through the middle of the condensates (x=0). At each z position we give the optical density of the atomic cloud, obtained directly as $\ln[I_0(x=0;z)/I(x=0;z)]$ where the ratio $I/I_0$ gives the normalized transmission profile.

The curves in Fig. 2 are results of mean-field calculations based on the Gross-Pitaevskii equation [25]. The trap potential used in the calculation is that of a three dimensional, asymmetric harmonic oscillator with $\omega_x=\omega_y=2050$ rad/s and $\omega_z=170$ rad/s. In the non-linear Gross-Pitaevskii equation we take the mean field potential to be $\frac{4\pi\hbar^2 a}{M}n$, where $n$ is the atomic density in the condensate, $M$ the sodium atomic mass, and $a$ the s-wave scattering length. We use the recent value of $a$=27.5 Å [26]. The Gross-Pitaevskii equation is solved numerically using the split-operator technique [27]. To obtain the theoretical profiles shown in Fig. 2, we propagate the probe beam through condensates, with atomic densities obtained from the mean field equation, and beyond through the imaging system. There is excellent agreement between the experimental images and the mean field calculations for 80,000, 350,000, and 850,000 atoms in the condensates, respectively (solid curves in Fig. 2).

In Figs 2-3 the arrows indicate the positions of the condensate surface (in the z direction) given by the Thomas-Fermi model [12], which can in each case be experimentally determined to better than 1%. In the Thomas-Fermi model, this size is $a_{TF} = \left(\frac{\hbar\lambda}{M\omega_z}\right)^{2/5}(15N_0 a)^{1/5}$, where $N_0$ is the number of atoms in the condensate and $\lambda = \frac{\omega_x}{\omega_z}$. The precision in the size determination allows the number of atoms in the condensate to be determined to within 5%. Alternatively, an independent measurement of $N_0$ allows for a high precision measurement of the scattering length. This philosophy was used for the measurements presented in Fig. 3, resulting in a scattering length for sodium atoms in a condensate of $a$=26.5Å±15%. This value agrees with the value of 27.5Å±10% determined in ref. 26 for a classical gas and with the value of 22.3Å±36% determined in ref. 28 from condensate release data. It should be noted that the statistical measurement uncertainty in our case is only 10%. The rest is connected with uncertainty in trap parameters and optical constants which future experimental and theoretical work will improve.

In Fig. 3a we show a 3D image of the optical density profile of a spherically symmetric condensate. The trapping potential was changed adiabatically after condensation to give a symmetric confining potential with ω = 87 rad/s. The number of atoms in the condensate was determined from a series of release experiments to 1.6 10$^6$ ±10%. The detuning of the probe beam was -43 MHz and the bias field was 38 gauss. In Fig. 3b we show the radial profile obtained by polar averaging of three data files like the one shown in Fig 3a. The solid curve in the figure is the column density calculated from the mean field equation.

In conclusion, we have measured the spatial density profiles of condensates directly in the magnetic bottle using the technique of near-resonant, absorption imaging. We have also introduced a new '4D' magnet which efficiently creates a highly confining potential for condensate formation. The imaging technique is sensitive to the detailed shape of the condensate wavefunction, in particular to the condensate surface. This allows us to determine condensate sizes to better than 1%. The scattering length for sodium atoms in a condensate is found to agree with the value obtained for a classical gas by photoassociative spectroscopy [26], supporting our believe that mean field theory is valid for these dilute gases. We expect that systematic improvements will bring the uncertainty in our determination of the scattering length below 10% which would make this method competitive with spectroscopic measurements. The ability to perform high-precision measurements near the condensate surface will be important in finite temperature studies of confined atom clouds where the condensate is expected to partially phase separate from the cloud of non-condensed atoms [29]. Near-resonant imaging should give detailed information about the interface and interaction between these components.

Acknowledgment.
We would like to thank Winfield Hill for his support regarding the electronics design, Cyrus Behroozi for assistance with calculations, and Don Rogers for machining essential parts of the setup. The research was supported by the Rowland Institute.

**Figures**

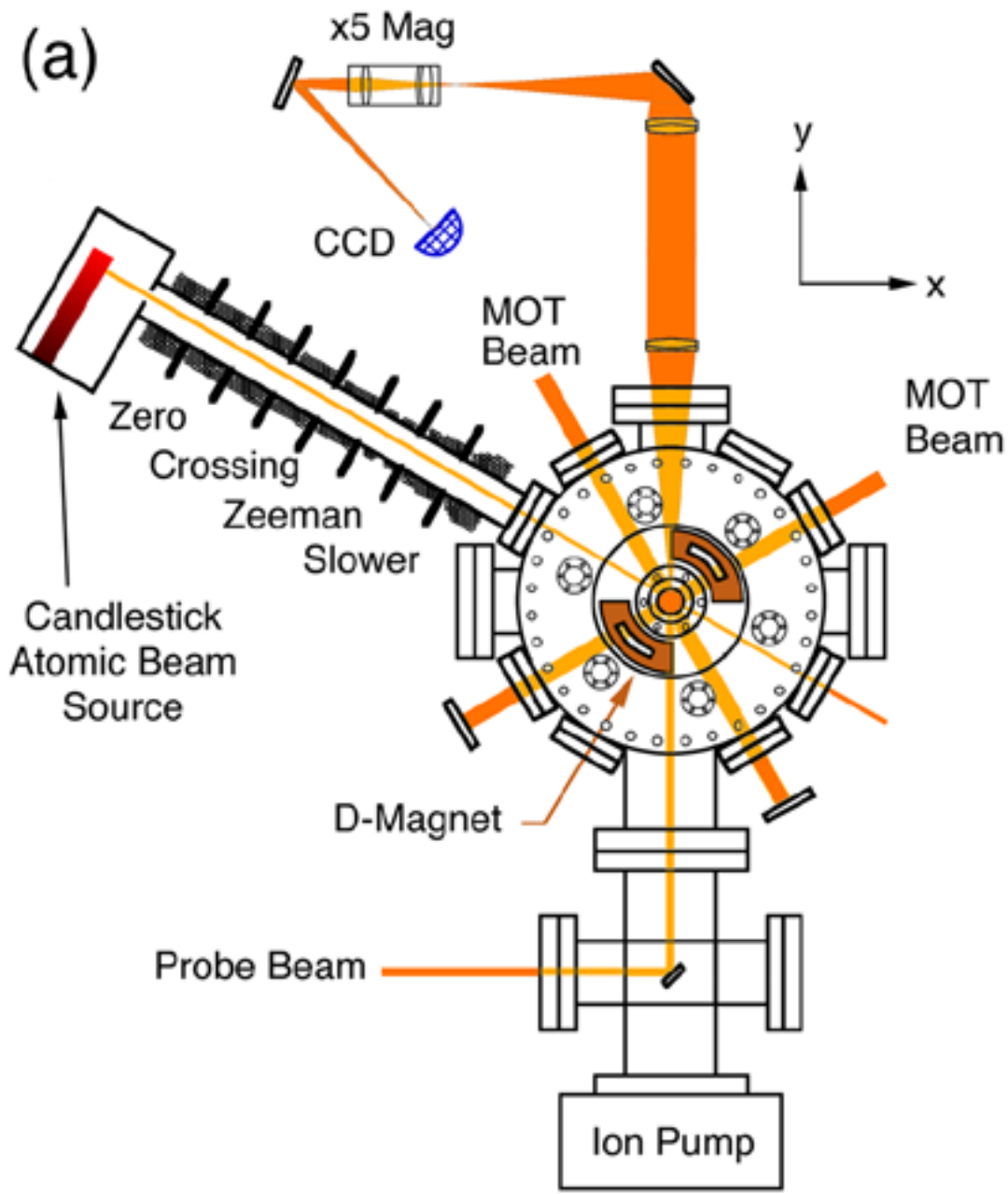

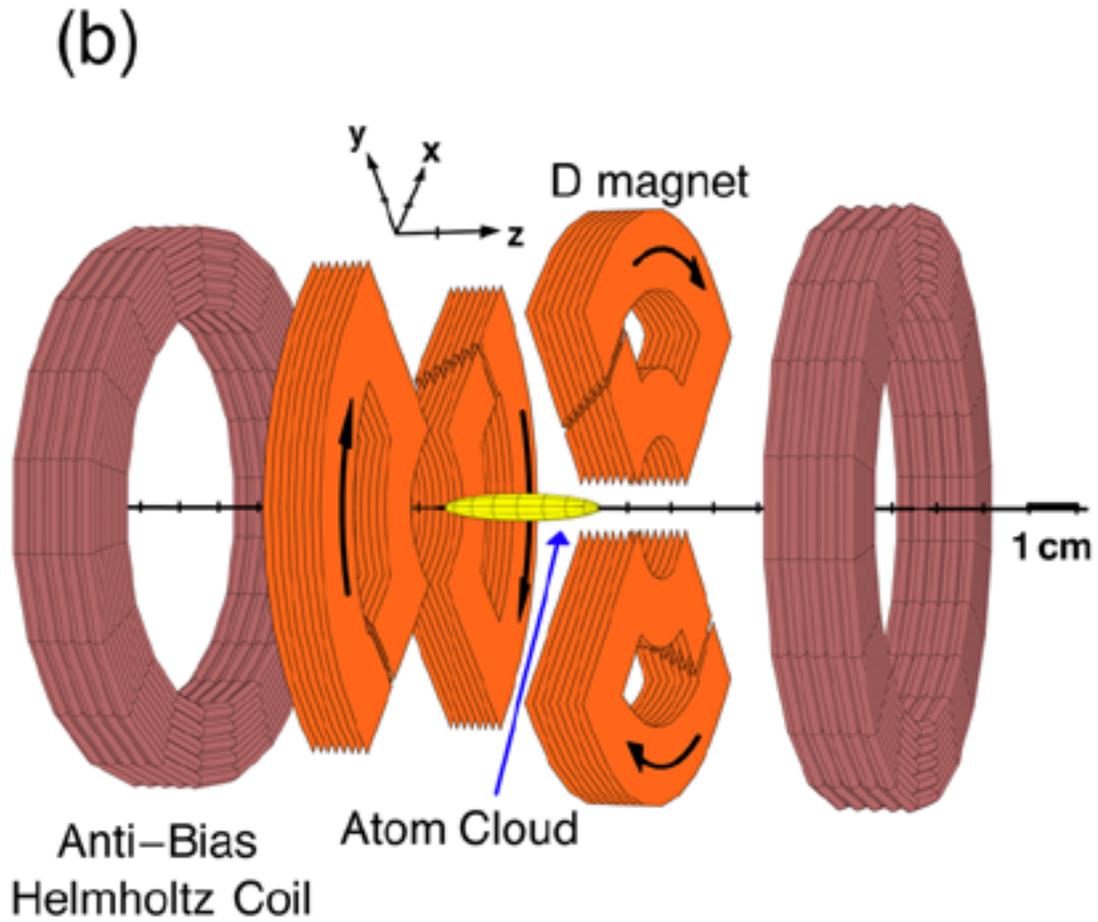

Figure 1. (a) Schematic layout of condensation apparatus viewed along the z direction. (b) Expanded view of the '4D' magnetic bottle used to evaporatively cool into the condensation regime.

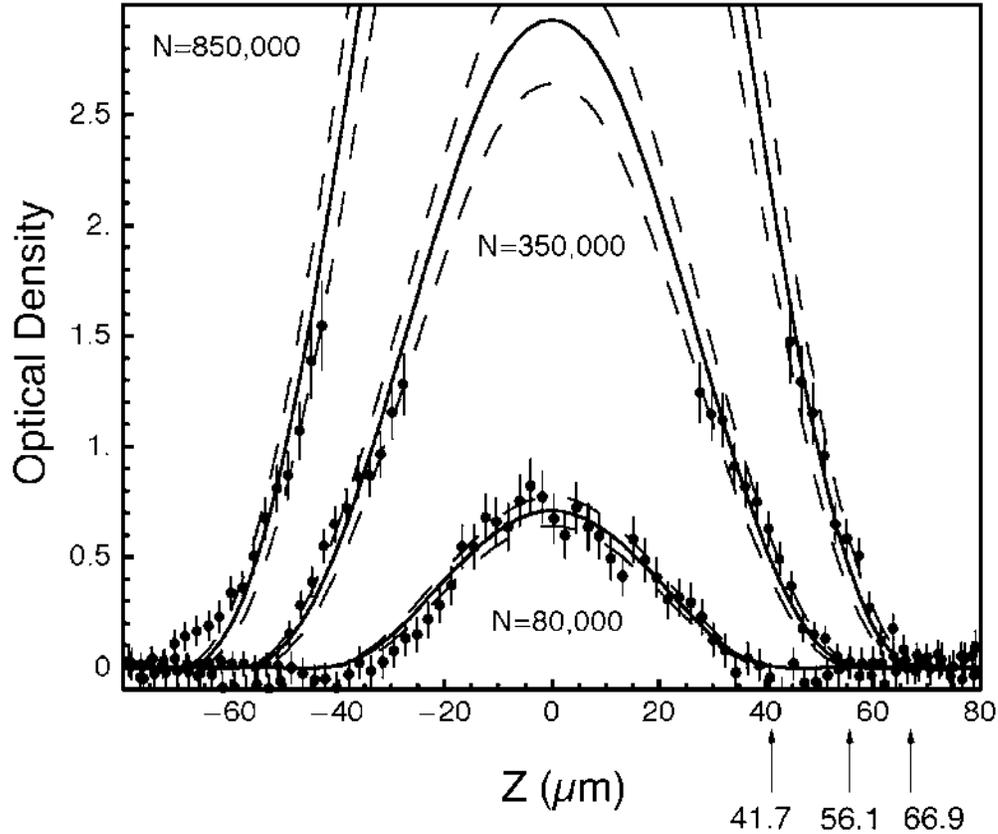

Figure 2. Near resonant images of column density profiles of sodium condensates. The solid curves result from mean-field calculations with $N_0$ = 80,000, 350,000, and 850,000 atoms in the condensates, respectively, after propagation through the optical system. These numbers correspond to peak atomic column densities of 442, 1072, and 1826 atoms/$\mu m^2$. The asymmetric harmonic oscillator potential has $\omega_x=\omega_y=2050$ rad/s and $\omega_z=170$ rad/s. Arrows give Thomas-Fermi sizes in the z direction. Error bars represent statistical errors. By comparing the theoretical curves to the data we obtain $\chi^2$ values of 0.71, 1.26, and 1.33. The upper/lower dashed curves in each case correspond to $N_0$ values shifted by ±10%.

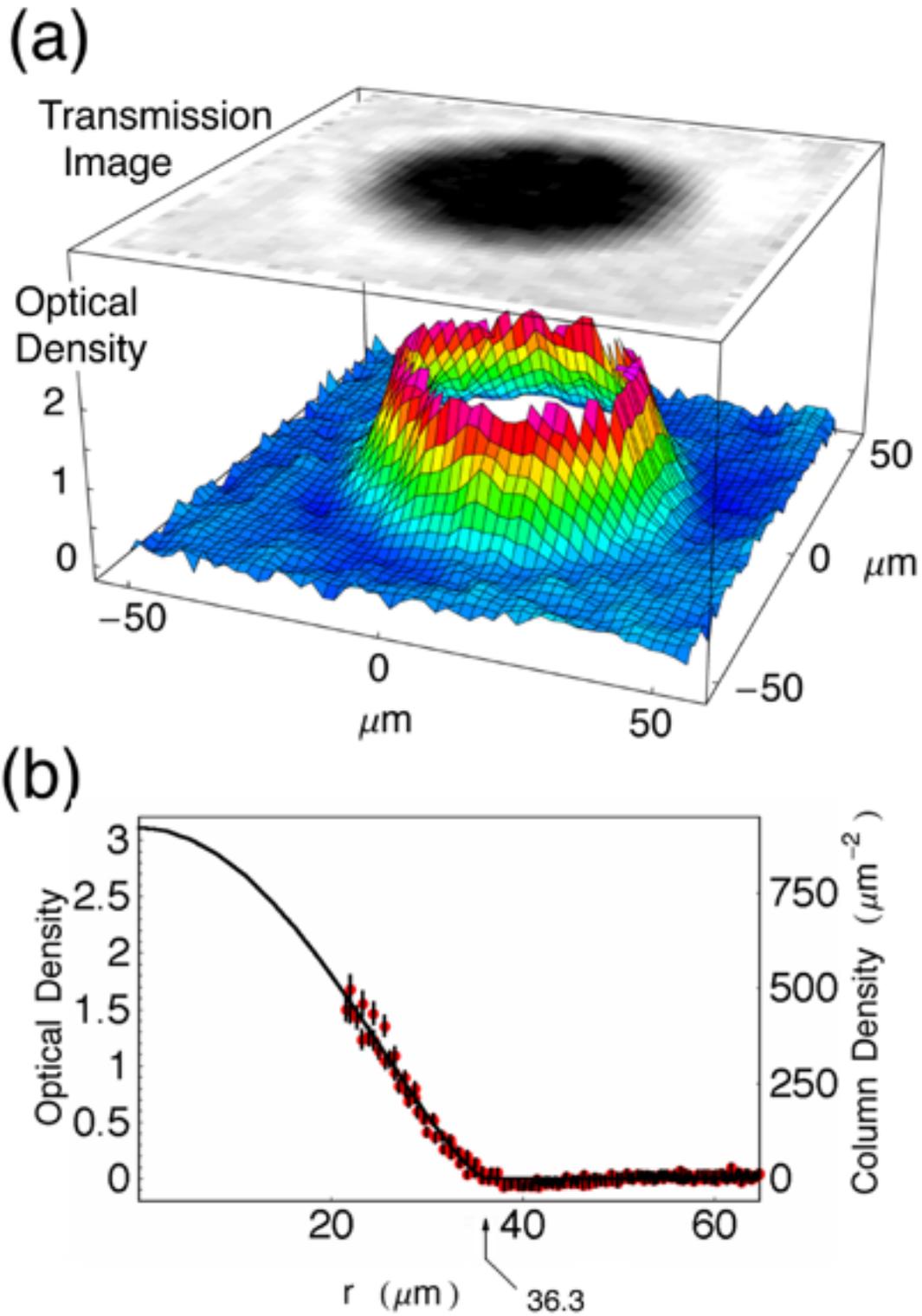

Figure 3. (a) 3D image of the condensate surface for a spherically symmetric condensate in the expanded trap with ω=87 rad/s. The gray-scale plot shows the normalized transmission profile. (b) Radial profile obtained by a polar (angular) average of the images of three different condensates. (This shows the high reproducibility of condensate formation in our system.) The solid curve is the result of a mean field calculation for 1.6 million atoms in the condensate as determined from release experiments. A comparison with the data gives $\chi^2$=1.23 [30].